\documentclass[aps,prc,letterpaper,11pt,twoside,tightenlines,nofootinbib,showpacs,preprint]{revtex4}
\usepackage{graphicx}
\usepackage[sort&compress]{natbib}
\usepackage{subfigure}
\usepackage{amsmath}
\usepackage{amsfonts}
\usepackage{cancel}
\usepackage{slashbox}
\usepackage{multirow}
\usepackage{pict2e}

\begin{document}

\arraycolsep1.5pt

\newcommand{\Ima}{\textrm{Im}}
\newcommand{\Rea}{\textrm{Re}}
\newcommand{\mev}{\textrm{ MeV}}
\newcommand{\be}{\begin{equation}}
\newcommand{\ee}{\end{equation}}
\newcommand{\ba}{\begin{eqnarray}}
\newcommand{\ea}{\end{eqnarray}}
\newcommand{\gev}{\textrm{ GeV}}
\newcommand{\nn}{{\nonumber}}
\newcommand{\dtres}{d^{\hspace{0.1mm} 3}\hspace{-0.5mm}}

\title{Chiral unitary approach to $\eta^\prime N$ scattering at low energies}

\author{E.~Oset}
\affiliation{
Departamento de F\'{\i}sica Te\'orica and IFIC, Centro Mixto Universidad de Valencia-CSIC,
Institutos de Investigaci\'on de Paterna, Aptdo. 22085, 46071 Valencia,
Spain}

\author{A. Ramos}
\affiliation{Departament d'Estructura i Constituents de la Mat\`eria and Institut de Ci\`encies del Cosmos. \\ Universitat de Barcelona,
Avda. Diagonal 647, 08028 Barcelona, Spain}

\date{\today}

 \begin{abstract}

We study the $\eta^\prime N$ interaction within a chiral unitary approach which includes $\pi N$, $\eta N$ and related pseudoscalar meson-baryon coupled channels. Since the SU(3) singlet does not contribute to the standard interaction and the $\eta^\prime$ is mostly a singlet, the resulting scattering amplitude is very small and inconsistent with experimental estimations of the
$\eta^\prime N$ scattering length. The additional consideration of vector meson-baryon states into the coupled channel scheme, via normal and anomalous couplings of pseudoscalar to vector mesons, enhances substantially the $\eta^\prime N$ amplitude. We also exploit the freedom of adding to the Lagrangian a new term, allowed by the symmetries of QCD, which  couples baryons to the singlet meson of SU(3). Adjusting the unknown strength to the $\eta^\prime N$ scattering length, we obtain predictions for the elastic $\eta^\prime N \to \eta^\prime N$ and inelastic $\eta^\prime N \to \eta  N$,  $\pi N$, $K\Lambda$, $K\Sigma$ cross sections at low $\eta^\prime$ energies, and discuss their significance.

\end{abstract}


\maketitle

\section{Introduction}
\label{Intro}

 The $\eta^\prime$ meson has interesting properties associated to the underlying QCD dynamics
of hadrons, in particular to the $U_A (1) $ axial vector anomaly \cite{kogut,weinberg,tooft,witten}. Being close to a singlet of SU(3), its interaction with nucleons is supposed to be weak compared for instance with the case of its partner, the $\eta$ meson. Yet, there are no theoretical calculations of the $\eta^\prime N$ interaction and only poor estimations of the value of the  $\eta^\prime N$ scattering length.  These estimations come from the study of the $pp \to pp \eta^\prime$ cross section near threshold at COSY  \cite{Moskal:2000gj,Moskal:2000pu}. By looking at the shape of this cross section it was concluded in \cite{Moskal:2000gj} that
$|{\rm Re}\,a_{\eta^\prime N}|< 0.8$ fm. A more refined analysis of the reaction, comparing the cross section with that of the $pp \to pp \pi ^0$ reaction to minimize the $pp$ final state interaction effects, concluded that the scattering length should be of the order of magnitude of that of the $\pi N$ and hence $|a_{\eta^\prime N}|\sim 0.1$ fm \cite{Moskal:2000pu}. This indicates a rather weak $\eta^\prime N$ interaction, much weaker than suggested in \cite{Baru:1999zg}
and \cite{Moskal:1998vm} based on earlier data that were much improved in
\cite{Moskal:2000gj}.

Even if only the approximate size of the scattering length is known, it is interesting to see what one can obtain in a theoretical approach. Also, the $\eta^\prime N$ interaction should have, even if weak, some repercussion in  $\eta^\prime$ production processes where one has a $\eta^\prime N$ pair in the final state, such as in the $\gamma p \to \eta^\prime p$  or $pp \to pp \eta^\prime$ reactions. The latter process has provided an experimental estimation of the $\eta^\prime N$ scattering length \cite{Moskal:2000gj,Moskal:2000pu}, as mentioned above, while
 the $\gamma p \to \eta^\prime p$ reaction has been investigated experimentally \cite{Struczinski:1975ik,Plotzke:1998ua,Barth:2001cb,Dugger:2005my,Crede:2009zzb} and theoretically \cite{Zhang:1995uha,Li:1996wj,Borasoy:2001pj,Zhao:2001kk,Chiang:2002vq,Sibirtsev:2003ng,Nakayama:2004ek,Nakayama:2005ts}. The theoretical approaches to $\eta^\prime$ photoproduction do not consider explicitly the $\eta^\prime N$ final state interaction, though it is implicitly assumed through the coupling of the $\eta^\prime N$ pair to some resonances. Yet, this is short of a refined unitary approach in coupled channels which would consider the $\eta^\prime N$ state together with its coupled meson-baryon states in the final system, rendering the problem into a many channel problem. The resonant models for the $\gamma p \to \eta^\prime p$ reactions do not provide information on the $\eta^\prime N$ scattering amplitude nor on how it could be tested experimentally.

 The aim of the present paper is to study explicitly the $\eta^\prime N$ interaction at low scattering energies. For this purpose, we have employed the chiral unitary approach to meson baryon interactions which has shown a remarkable success in the study of this type of processes  \cite{Kaiser:1995cy,angels,ollerulf,carmenjuan,hyodo,bennhold,inoue,nsd,juanenrique,Nacher:1999vg,Nieves:2001wt}. A necessary ingredient in the calculation is the mixing of the SU(3) singlet ($\eta_1$) and octet ($\eta_8$) states present in the physical $\eta$ and $\eta^\prime$ mesons. The amount of singlet and octet components of the two particles is a problem that has attracted much attention \cite{Gilman:1987ax,Bramon:1989kk,Schechter:1992iz,Leutwyler:1996sa,Feldmann:1998vh,Bramon:1997va,Mathieu:2010ss,ambrosino}, and there are good recent determinations of this mixing angle.
 
We start our study by considering, in the coupled channel scheme, the basis of pseudoscalar meson-baryon states ($PB$) containing $\pi N$, $\eta N$, $\eta^\prime N$, $K \Lambda$ and $K \Sigma$. The resulting $\eta^\prime N$ scattering amplitude is very small since, given the structure of the standard Lagrangian employed, it is obtained from the octet component of the $\eta^\prime$ meson. However, the energy region around the $\eta^\prime N$ threshold lies very close to the position of a resonance with the quantum numbers of $\eta^\prime N$, generated dynamically from the vector meson-baryon ($VB$) interaction in s-wave \cite{angelsvec}. This state will certainly have an influence on the $\eta^\prime N$ interaction, worth to be explored. For this purpose, we have implemented in our formalism the relevant $VB$ channels that generate this resonance, allowing them to couple to the $PB$ components. This produces a substantial enhancement of the $\eta^\prime N$ amplitude, yet insufficient to account for the empirical $\eta^\prime N$ scattering length. Finally, we note that the symmetries of QCD also allow for a term in the Lagrangian coupling the baryons to the singlet component of the pseudoscalar meson \cite{borasinglet,Kawarabayashi:1980uh}. We exploit this freedom and, adjusting the unknown strength of this new term to the empirical value of the $\eta^\prime N$ scattering length, we obtain the elastic $\eta^\prime N$ cross section and transitions of $\eta^\prime N$ to the different coupled channels. We find that, while the elastic $\eta^\prime N$ cross section is very sensitive to the free parameter of the theory, the inelastic cross sections are rather stable and constitute a genuine prediction of the model.

\section{The \mbox{\boldmath$\eta^\prime$} N interaction from a coupled pseudoscalar-baryon basis.}

  We obtain the $\eta^\prime N$ interaction using the chiral unitary approach in coupled channels of
\cite{Kaiser:1995cy,angels,ollerulf,carmenjuan,hyodo,bennhold,inoue,nsd,juanenrique,Nacher:1999vg,Nieves:2001wt}. We essentially follow here the procedure of \cite{angels}, updated in \cite{bennhold} to energies not so close to threshold, and, particularly, the model of Ref.~\cite{inoue}, which studies the interaction of $\pi N, \eta N$ and related coupled channels. The meson-baryon scattering dynamics is taken from the lowest order chiral Lagrangian reduced to the two meson fields needed in the process,
\begin{equation}
{\cal L}_1^{(B)} = < \bar{B} i \gamma^{\mu} \frac{1}{4 f^2}
[(\Phi \partial_{\mu} \Phi - \partial_{\mu} \Phi \Phi) B
- B (\Phi \partial_{\mu} \Phi - \partial_{\mu} \Phi \Phi)] >
\label{lagran}
\end{equation}
where $f$ is the pion decay constant
and $\Phi$, $B$ are
the SU(3) matrices for the mesons and the baryons:
\begin{equation}
\Phi =
\left(
\begin{array}{ccc}
\frac{1}{\sqrt{2}} \pi^0 + \frac{1}{\sqrt{6}} \eta_8 & \pi^+ & K^+ \\
\pi^- & - \frac{1}{\sqrt{2}} \pi^0 + \frac{1}{\sqrt{6}} \eta_8 & K^0 \\
K^- & \bar{K}^0 & - \frac{2}{\sqrt{6}} \eta_8
\end{array}
\right) \ ,
\label{eq:phi}
\end{equation}
and
\begin{equation}
B =
\left(
\begin{array}{ccc}
\frac{1}{\sqrt{2}} \Sigma^0 + \frac{1}{\sqrt{6}} \Lambda &
\Sigma^+ & p \\
\Sigma^- & - \frac{1}{\sqrt{2}} \Sigma^0 + \frac{1}{\sqrt{6}} \Lambda & n \\
\Xi^- & \Xi^0 & - \frac{2}{\sqrt{6}} \Lambda
\end{array}
\right) \ .
\label{eq:B}
\end{equation}
The field $\eta_8$ in Eq. (\ref{eq:phi}) is a member of the pseudoscalar Goldstone boson octet. In order to deal with the physical $\eta$ and $\eta^\prime$ mesons, we have to introduce the singlet $\eta_1$, which is easily accomplished by adding the diagonal matrix,
${\rm diag}(\eta_1/\sqrt{3},\eta_1/\sqrt{3},\eta_1/\sqrt{3})$, to the matrix $\Phi$ of Eq.~(\ref{eq:phi}).

 The physical $\eta,\eta^\prime$ meson fields are related to the SU(3) singlet and octet fields, $\eta_1$ and $\eta_8$, via the relations:
   \begin{eqnarray}
\eta &=& \cos\theta_P\, \eta_8 - \sin\theta_P\, \eta_1   \nonumber \\
\eta' &=& \sin\theta_P\, \eta_8 + \cos\theta_P\, \eta_1  \ ,
\label{eq:mezclas}
\end{eqnarray}
where $\theta_P$ is the  $\eta_1-\eta_8$ mixing angle, for which we take the value $\theta_P=-14.34^\circ$ reported in
\cite{ambrosino}, which was determined phenomenologically from fits to the world data. This corresponds to a mixing angle of strange to nonstrange quarks of $40.4^\circ$. The former equation establishes the $\eta$ as being largely an octet while the $\eta^\prime$ is mostly a singlet.

Note that the particular structure of the standard Lagrangian of
Eq.~(\ref{lagran}),
$\Phi \partial_{\mu} \Phi - \partial_{\mu} \Phi \Phi$, makes the contribution of the singlet vanish. In other words, within this chiral formalism, a pure singlet $\eta^\prime$ state would not interact with the nucleons.
This is at the root of why, as we will see, the $\eta^\prime N$ interaction is so weak, since the magnitude of this interaction is tied to the small octet content of the $\eta^\prime$.

From Eq. (\ref{lagran}) we obtain a kernel, or transition potential $V$, in s-wave:
   \begin{equation}
V_{i j} = - C_{i j} \frac{1}{4 f_i f_j}(2\sqrt{s} - M_{i}-M_{j})
\left(\frac{M_{i}+E_i}{2M_{i}}\right)^{1/2} \left(\frac{M_{j}+E_j}{2M_{j}}
\right)^{1/2}\, ,
\label{eq:ampl2}
\end{equation}
where $\sqrt{s}$ is the total energy in the center-of-mass frame and $M_i, E_i$ ($M_j,E_j$) stand for the mass, energy of the incoming (outgoing) baryon. The coupled isospin $I=1/2$ channels of the present problem are $\pi N$, $\eta N$, $\eta^\prime N$, $K \Lambda$ and $K \Sigma$.  The corresponding SU(3) coupling coefficients $C_{i j}$ are evaluated in \cite{inoue}. However, since only the octet component of the fields participates in the interaction, the couplings involving $\eta$ or $\eta^\prime$ mesons are those of the $\eta_8$ but multiplied by $\cos\theta_P$ for each $\eta$ state and by  $\sin\theta_P$ for each $\eta^\prime$ state in the channels studied.  As in Ref.~\cite{inoue}, we take different weak decay constants for each meson involved in the transition. We use the values $f_\pi=93$ MeV, $f_K=1.22 f_\pi$ and $f_\eta=1.3 f_\pi$ taken from chiral perturbation theory \cite{Gasser:1984gg}. With the potential of Eq.~(\ref{eq:ampl2}) we solve the Bethe-Salpeter equation in its on shell factorized form \cite{nsd,ollerulf,juanenrique}:
\begin{equation}
T = [1 - V \, G]^{-1}\, V
\label{eq:bs1}
\end{equation}
with $V$ the matrix of Eq.~(\ref{eq:ampl2}) evaluated on shell, and $G$ the loop function for the intermediate meson baryon propagators. Following  \cite{ollerulf}, we employ dimensional regularization and replace the divergent pieces by subtraction constants
 $a_l(\mu)$, which depend on the chosen energy scale $\mu$, obtaining:
\begin{eqnarray}
G_{l} &=& i 2 M_l \int \frac{d^4 q}{(2 \pi)^4} \,
\frac{1}{(P-q)^2 - M_l^2 + i \epsilon} \, \frac{1}{q^2 - m^2_l + i
\epsilon}  \nonumber \\ &=& \frac{2 M_l}{16 \pi^2} \left\{ a_l(\mu) + \ln
\frac{M_l^2}{\mu^2} + \frac{m_l^2-M_l^2 + s}{2s} \ln \frac{m_l^2}{M_l^2} +
\right. \nonumber \\ & &  \phantom{\frac{2 M}{16 \pi^2}} +
\frac{\bar{q}_l}{\sqrt{s}}
\left[
\ln(s-(M_l^2-m_l^2)+2\bar{q}_l\sqrt{s})+
\ln(s+(M_l^2-m_l^2)+2\bar{q}_l\sqrt{s}) \right. \nonumber  \\
& & \left. \phantom{\frac{2 M}{16 \pi^2} +
\frac{\bar{q}_l}{\sqrt{s}}}
\left. \hspace*{-0.3cm}- \ln(-s+(M_l^2-m_l^2)+2\bar{q}_l\sqrt{s})-
\ln(-s-(M_l^2-m_l^2)+2\bar{q}_l\sqrt{s}) \right]
\right\} \ ,
\label{eq:gpropdr}
\end{eqnarray}
where $\bar{q}_l$ is the momentum of the meson and baryon of channel $l$ in the c.m. frame. Usually, given a value of $\mu$, the subtraction constants are left as free parameters that are fitted to reproduce the available scattering observables in a certain energy region. The simplest model used in Ref.~\cite{inoue} for studying $\pi N$ scattering and related channels in the strangeness $S=0$ sector adopted
the following values:
\begin{eqnarray}
 &&\mu=1200~{\rm MeV}\nonumber \\
 &&a_{\pi N}=2.0,~~a_{\eta_8 N}=
0.2,~~a_{K \Lambda}=1.6,~~a_{K\Sigma}=-2.8 \ .
\label{eq:inoue_set}
\end{eqnarray}
This set of parameters leads to a qualitative agreement with the s-wave amplitudes in the energy range from threshold to 1600 MeV. In particular, it
generates the $N(1535)$ $S_{11}$ resonance, which is common to all the unitary approaches \cite{Kaiser:1995cy,Nacher:1999vg,Nieves:2001wt}. Although the model of Ref.~\cite{inoue} was improved with the inclusion of the $\pi\pi N$ channel, which was shown to have an important repercussion especially in the isospin 3/2 sector, for the purpose of the present work, aiming at obtaining the isospin $I=1/2$ $\eta^\prime N$ amplitude, the simplified coupled two-body channel model is sufficient. We note that the imaginary part of the $\eta N$ scattering length is constrained by unitarity and a lower bound for it comes from the the cross section $\pi N \to \eta N$, namely ${\rm Im}\, a_{\eta N} \ge 0.172 \pm 0.009$ fm \cite{Arndt:2005dg}. Since the incorporation of the $\eta_1-\eta_8$ mixing modifies slightly the $\eta N$ interaction, the model parameters have also been fine tuned in the present work to the values:
\begin{eqnarray}
 &&\mu=1200~{\rm MeV}\nonumber \\
 &&a_{\pi N}=2.0,~~a_{\eta N}=0.9,~~a_{\eta^\prime N}=0.9,
~~a_{K \Lambda}=2.2,~~a_{K\Sigma}=-3.1 \ ,
\label{eq:our_set}
\end{eqnarray}
in order to comply with the unitarity constraint for $\eta N$ scattering.

 In \cite{borasinglet} it is shown that there is another type of interaction, coupling the singlet component of the meson field to the baryons, and written as $\lambda_1 2 M_B \eta_1 \eta_1 \langle \bar B B \rangle$ in the static limit, that can contribute to the $\eta^\prime N$ interaction with a strength governed by $\lambda_1$. Similar developments were made in \cite{Kawarabayashi:1980uh}. This term was justified to be small in Ref.~\cite{borasinglet} and was neglected. In the present work, we explore the influence of this term on the $\eta^\prime N$ scattering observables. It will only affect transitions involving $\eta$ and $\eta^\prime$ mesons which have a singlet component. Taking into account Eq.~(\ref{eq:mezclas}), we can write the singlet transition potentials between $\eta N$ and $\eta^\prime N$ states as
 \begin{equation}
V^{(1)}_{\eta N, \eta N}= C \sin^2 \theta_P;~~~~~~ V^{(1)}_{\eta N, \eta^\prime N}= -C \sin \theta_P \cos \theta_P ~~~~~~ V^{(1)}_{\eta^\prime N, \eta^\prime N}= C \cos^2 \theta_P \ ,
\label{eq:singletcoup}
\end{equation}
where the singlet strength has been parametrized in the form:
\begin{equation}
C=\frac{\alpha}{4 f_\pi^2} 2 m_{\eta^\prime}\frac{E_B+E_{B'}}{2M_N}\ ,
\end{equation}
with $\alpha$ an unknown parameter. We have chosen this parametrization to allow a more straightforward comparison with the terms of the Lagrangian coming from the octet components of the fields given in Eq. (\ref{eq:ampl2}).

\section{Coupling to vector-baryon channels}
The interaction of vector mesons and baryons was studied in \cite{angelsvec}, for the baryons of the octet of the proton and vector mesons of the nonet of the $\rho$, and in \cite{Sarkar:2009kx}, for the nonet of vector mesons and the decuplet of baryons of the $\Delta(1232)$. In both cases, many dynamically generated resonances where obtained, some of which compared favorably with particles in the Particle Data Book (PDG) \cite{Nakamura:2010zzi}. The threshold for $\eta^\prime N$ is around 1900 MeV, very close to the location of dynamically generated states in Ref.~\cite{angelsvec} with quantum numbers $J^P=1/2^-$, as well as their degenerate $J^P=3/2^-$ partners, which could correspond to some experimental resonances reported around this region. 

In order to incorporate the effect of the resonances in the $\eta^\prime N$ interaction, we must establish the transition from the $PB$ to the $VB$ channels which generate these states by multiple scattering. We are particularly interested in the $1/2^-$ resonance appearing around 1970 MeV in \cite{angelsvec}. It couples mostly to the $K^* \Lambda$ and $K^* \Sigma$ channels and decouples from the $\rho N$ channel. The coupling to the $\omega N$ and $\phi N$ channels is very weak and, in addition, the $\phi N$  coupling is further OZI suppressed in the $\eta^\prime N \to \phi N$ transition potential. This leaves only $K^* \Lambda$ and $K^* \Sigma$ as the relevant $VB$ channels to be considered in our study.

\begin{figure}[ht]
\begin{center}
\includegraphics[width=0.7\textwidth]{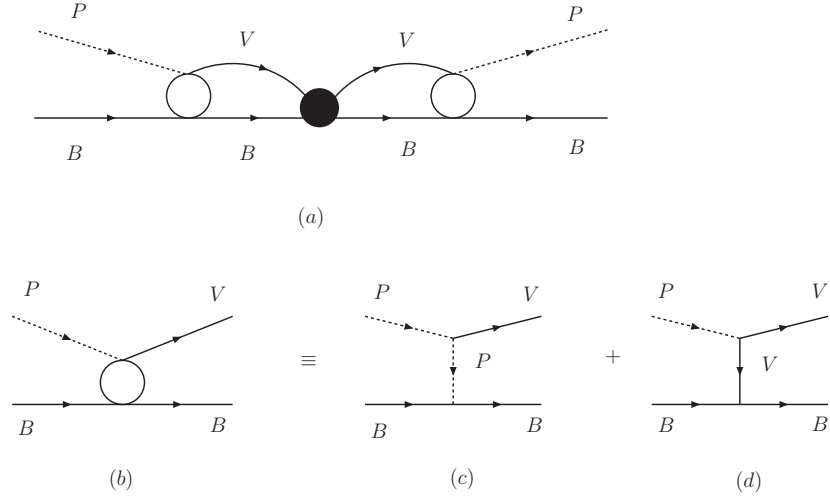}
\end{center}
\caption{Diagrams considered to include the intermediate vector mesons-baryon states with their interaction.}
\label{fig:vectorloop}
\end{figure}

 Technically, we have implemented the $PB-VB$ coupling by still working within a basis of coupled $PB$ states but including explicitly, as part of a $PB-PB$ potential, the transition of the $PB$ to the $VB$ channels, the later ones interacting among themselves to produce the resonance, as depicted schematically by the diagram of
 Fig.~\ref{fig:vectorloop}(a).

 The mechanisms for the $PB-VB$ transition have been reported in Ref.~\cite{Oset:2011ui}, but here we will also have to include a contribution from anomalous terms, given the fact that the standard contributions of  \cite{Oset:2011ui} will be suppressed by $\sin\theta_P$.
More specifically, the $PB$ to $VB$ transition is done via the normal pseudoscalar exchange involving the $V \to PP$ decay vertex [Fig.~\ref{fig:vectorloop}(c)], and via the anomalous term with vector exchange involving the  $V \to VP$ decay vertex [Fig.~\ref{fig:vectorloop}(d)]. The  Lagrangian for the normal coupling is given by
 \begin{equation}
{\cal L}_{VPP}= -ig \langle [
P,\partial_{\nu}P]V^{\nu}\rangle \ ,
\label{eq:lagrVpp}
\end{equation}
where $P$ is the SU(3) matrix of the pseudoscalar fields of Eq.~(\ref{eq:phi}) and
$V_\mu$ is the SU(3)
matrix of the vector mesons of the octet of the $\rho$
\begin{equation}
V_\mu=\left(
\begin{array}{ccc}
\frac{\rho^0}{\sqrt{2}}+\frac{\omega}{\sqrt{2}}&\rho^+& K^{*+}\\
\rho^-& -\frac{\rho^0}{\sqrt{2}}+\frac{\omega}{\sqrt{2}}&K^{*0}\\
K^{*-}& \bar{K}^{*0}&\phi\\
\end{array}
\right)_\mu \ .
\label{Vmu}
\end{equation}
For the anomalous coupling we use the VMD Lagrangians involving
the vertex $VVP$ 
of \cite{Bramon:1992kr}
\begin{equation}
{\cal L}_{VVP} = \frac{G}{\sqrt{2}}\epsilon^{\mu \nu \alpha \beta}\langle
\partial_{\mu} V_{\nu} \partial_{\alpha} V_{\beta} P \rangle \ ,
\label{eq:lagranom}
\end{equation}
where $G=\displaystyle\frac{3{g^\prime}^2}{4\pi^2f}$,
$g^\prime=-\displaystyle\frac{G_VM_{\rho}}{\sqrt{2}f^2}$ \cite{Bramon:1992kr}
and $f=f_\pi=93\,$MeV,
 with $G_V$ being the
coupling of $\rho$ to $\pi\pi$ in the normalization of
\cite{Ecker:1988te}, $G_V=53$ MeV.

For the baryon vertices, in Fig.~\ref{fig:vectorloop}(c) we need the Yukawa $PBB$ Lagrangian given by
\begin{equation}
{\cal L}_{PBB}=-\sqrt 2 \frac{D+F}{2f} \langle\bar{B}\gamma^{\mu}\gamma_5 \partial_{\mu}\phi B \rangle -
\sqrt 2 \frac{D-F}{2f} \langle\bar{B}\gamma^{\mu}\gamma_5 B\partial_{\mu}\phi \rangle \ ,
\label{eq:lagrpbb}
\end{equation}
with $D=0.795$, $F=0.465$ taken from \cite{boradf}, while in Fig.~\ref{fig:vectorloop}(d)
we use the Lagrangian for the coupling of vector mesons to
the baryon octet given by
\cite{Klingl:1997kf,Palomar:2002hk} \footnote{Correcting a misprint in
\cite{Klingl:1997kf}}
\begin{equation}
{\cal L}_{BBV} =
g\left( \langle \bar{B}\gamma_{\mu}[V^{\mu},B]\rangle +
\langle \bar{B}\gamma_{\mu}B \rangle \langle V^{\mu}\rangle \right) \ .
\label{lagr82}
\end{equation}

In the evaluation of the transition potentials,  we follow the approach of \cite{angelsvec} and neglect the three momenta of the vector mesons with respect to their mass, which allows one to neglect the $\epsilon^0$ polarization component coming from the Lagrangians of Eqs.~(\ref{eq:lagrVpp}), (\ref{eq:lagranom}) and (\ref{lagr82}) and keep only the spatial components. Along the same lines, the driving term  $\gamma^{\mu}\gamma_5$ of the Lagrangian of Eq.~(\ref{eq:lagrpbb}) is approximated by  $\vec{\sigma} \vec{q}$ in our calculations.

Denoting by $a,b$ any two of the $PB$ channels ($\pi N, \eta N, \eta^\prime N, K \Lambda, K \Sigma$) we finally obtain the $PB \to PB$ transition corresponding to the diagram of Fig.~\ref{fig:vectorloop}(a) as:
 \begin{equation}
 \delta V_{ab}= \sum_{kl}\tilde{V}_{ak} T_{kl} \tilde{V}_{bl} \ ,
 \label{eq:v_int}
 \end{equation}
where the indices $k,l$ denote the $VB$ channels, and $T_{kl}$ are the $VB \to VB$ amplitudes evaluated in \cite{angelsvec}. The vertex $\tilde{V}_{ak}$ is given by:
\begin{equation}
\tilde{V}_{ak}=\int\frac{d^3q}{(2\pi)^3}\tilde{C}_{ak}(\vec{q\,}) \frac{1}{2\omega_{K^*}(\vec{q\,})}\frac{M_k}{E_k(\vec{q\,})}\frac{1}{\sqrt{s}-\omega_{K^*}(\vec{q\,})-
E_k(\vec{q\,})+i\varepsilon}F^2(\vec{q\,}) \ ,
\label{eq:vertex}
\end{equation}
where $\omega_{K^*}(\vec{q\,})=\sqrt{m_{K^*}^2+\vec{q}\,^2}$ is the energy of the $K^*$, with $m_{K^*}$ being the $K^*$ mass, and $M_k$, $E_k$ stand for the mass and energy of the baryon in the $VB$ channel ($K^*\Lambda$, $K^*\Sigma$). The functions $\tilde{C}_{ak}(\vec{q\,})$, calculated from the Lagrangians of Eqs.~(\ref{eq:lagrVpp}) to (\ref{lagr82}) and the appropriate meson propagators, are given in Table~\ref{tab:vertices}, where
\begin{equation}
H_K(\vec{q}\,)=\frac{\vec{q}\,^2}{\vec{q}\,^2 + m_K^2},~~~~
H_\pi(\vec{q}\,)=\frac{\vec{q}\,^2}{\vec{q}\,^2 + m_\pi^2},~~~~
H_\eta(\vec{q}\,)=\frac{\vec{q}\,^2}{\vec{q}\,^2 + m_\eta^2} \ ,
\end{equation}
with $m_K$, $m_\pi$ and $m_\eta$, the masses of the kaon, pion and eta mesons. The integral of Eq.~(\ref{eq:vertex}) is regularized by a form factor:
\begin{equation}
F(\vec{q}\,)=\frac{\Lambda^2}{\Lambda^2+\vec{q}\,^2}  \ ,
\end{equation}
typical of a Yukawa $PPB$ coupling, where a value $\Lambda=1200$ MeV has been taken.

\begin{table}[htbp]
    \setlength{\tabcolsep}{0.1cm}
\begin{center}
\begin{tabular}{l|cc}
\backslashbox{$a$}{$k$} & $K^* \Lambda$ & $K^* \Sigma$ \\
\hline
\\
 $\pi N$ & $g \displaystyle\frac{1}{\sqrt{6}}\displaystyle\frac{D+3F}{2f} H_K(\vec{q}\,)$ &
 $g \displaystyle\frac{1}{\sqrt{6}}\displaystyle\frac{D-F}{2f} H_K(\vec{q}\,)$ \\
\\
 $\eta N$ &  $-g  \displaystyle\frac{1}{\sqrt{6}}\displaystyle\frac{D+3F}{2f} \cos{\theta_P} H_K(\vec{q}\,)$ &

 $g \displaystyle\frac{3}{\sqrt{6}}\displaystyle\frac{D-F}{2f}  \cos{\theta_P} H_K(\vec{q}\,)$ \\
\\
 $\eta^\prime N$ & $-g  \displaystyle\frac{1}{\sqrt{6}}\displaystyle\frac{D+3F}{2f} \sin{\theta_P} H_K(\vec{q}\,)$ &
 $g \displaystyle\frac{3}{\sqrt{6}}\displaystyle\frac{D-F}{2f}  \sin{\theta_P} H_K(\vec{q}\,)$ \\
\\
 $K \Lambda$ &  $g \displaystyle\frac{1}{\sqrt{6}}\displaystyle\frac{2D}{2f} \cos{\theta_P} H_\eta(\vec{q}\,)$ & $-g \displaystyle\frac{2}{\sqrt{6}}\displaystyle\frac{D}{2f} H_\pi(\vec{q}\,)$ \\
\\
 $K\Sigma$ & $-g \displaystyle\frac{2}{\sqrt{6}}\displaystyle\frac{D}{2f} H_\pi(\vec{q}\,)$ &
 $g \displaystyle\frac{2}{\sqrt{6}}\displaystyle\frac{2F}{2f} H_\pi(\vec{q}\,) -
 g \displaystyle\frac{1}{\sqrt{6}}\displaystyle\frac{2D}{2f} \cos{\theta_P} H_\eta(\vec{q}\,)$
 \\
 \\
\hline
\end{tabular}
\end{center}
\caption{Functions $\tilde{C}_{ak}(\vec{q\,})$ involved in the evaluation of the $PB$ to $VB$ transition vertices of Eq.~(\ref{eq:vertex}). \label{tab:vertices}}
\end{table}

For the case of the $\eta^\prime N \to k$ transition we also take into account the anomalous vertex, and have
 \begin{equation}
 \tilde{V}_{\eta^\prime N, k} = \tilde{V}^{\rm norm}_{\eta^\prime N, k}+\tilde{V}^{\rm anom}_{\eta^\prime N, k}
 \end{equation}
 where  $\tilde{V}^{\rm norm}_{\eta^\prime N, k}$ is the function given above in Eq.~(\ref{eq:vertex}) and  $\tilde{V}^{\rm anom}_{\eta^\prime N, k}$ is
 also obtained from Eq.~(\ref{eq:vertex}), replacing $\tilde{C}_{ak}(\vec{q\,})$ by
 \begin{equation}
 \tilde{C}_{\eta^\prime N, k}^{\rm anom}(\vec{q\,})=
 \frac{1}{\sqrt{3}}\frac{3}{\sqrt{6}}g \frac{G}{\sqrt{2}}\frac{m_{K^*}}{M_k} \left(\frac{2\cos{\theta_P}}{\sqrt{3}}-\frac{\sin{\theta_P}}{\sqrt{6}}\right) \frac{\vec{q}\,^2}{\vec{q}\,^2 + m_{K^*}^2} \ .
 \label{eq:v_anom}
 \end{equation}

 For the octet mesons and the octet-dominated $\eta$ meson, the normal vertex of Eq.~(\ref{eq:lagrVpp}) is much larger than the anomalous one coming from Eq.~(\ref{eq:lagranom}), which can be safely neglected. However, in the case of the $\eta^\prime$, mostly a singlet, the anomalous contribution becomes very important and numerically larger than the ordinary one, which is proportional to $\sin \theta_P$
 as we can see from Table~\ref{tab:vertices}. The reason is that, as mentioned before, there is no contribution
 of the singlet component with the structure of the Lagrangian of  Eq. (\ref{eq:lagrVpp}), while, on the other hand, $\tilde{V}^{\rm anom}_{\eta^\prime N, k}$ receives a contribution from the singlet of pseudoscalar mesons.  Therefore, the normal vertex is bigger than the anomalous one for the octet mesons but, in the case of the $\eta^\prime$, mostly a singlet, the anomalous contribution becomes dominant.

\begin{figure}[ht]
\begin{center}
\includegraphics[width=0.4\textwidth]{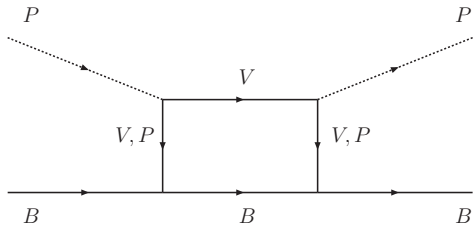}
\end{center}
\caption{Box diagrams coupling the $PB$ states to the intermediate $VB$ states without their interaction.}
\label{fig:Box}
\end{figure}

   We must also evaluate the contribution of the box diagram containing vector-baryon intermediate states without interaction. This is technically different than the case with interaction since the loop contains four p-wave vertices, as we see in Fig.~\ref{fig:Box}.

   Using the Lagrangians described above, the evaluation proceeds analogously and we have
 \begin{equation}
 \delta V_{ab}^{\rm box}=\sum_{k}\tilde{G}^{(ab)}_k \ ,
  \label{eq:v_box}
 \end{equation}
   where
   \begin{equation}
   \tilde{G}^{(ab)}_k = \int\frac{d^3q}{(2\pi)^3} \tilde{C}_{ak}(\vec{q\,})
   \tilde{C}_{bk}(\vec{q\,})  \frac{1}{2\omega_{K^*}(\vec{q\,})}\frac{M_k}{E_k(\vec{q\,})}\frac{1}{\sqrt{s}-\omega_{K^*}(\vec{q\,})-
   E_k(\vec{q\,})+i\varepsilon} F^2(\vec{q\,}) \ .
   \end{equation}
In the case of $a$ or $b$ being the $\eta^\prime N$ channel, the corresponding    $\tilde{C}_{ak}(\vec{q\,})$ or $\tilde{C}_{bk}(\vec{q\,})$ function must include, in addition to the normal term of Table~\ref{tab:vertices}, the anomalous contribution given by  Eq.~(\ref{eq:v_anom}).

The complete $PB-PB$ kernel, to be used in the Bethe-Salpeter equation of Eq.~(\ref{eq:bs1}), will then contain the standard Weinberg-Tomozawa (WT) contributions of Eq.~(\ref{eq:ampl2}), including the singlet terms of Eq.~(\ref{eq:singletcoup}) for transitions between $\eta N$ and $\eta^\prime N$ states, plus the additional contributions coming from the coupling to $VB$ states, namely the box $\delta V_{ab}^{\rm box}$ term of Eq.~(\ref{eq:v_box}) and the interacting $\delta V_{ab}$ term of Eq.~(\ref{eq:v_int}).

\section{Results}

   In the first place, we show in Table~\ref{tab:scat} results for the $\eta^\prime p$ scattering length, $a_{\eta^\prime p}$, calculated from the scattering amplitude $t_{\eta^\prime p \to \eta^\prime p }$ at threshold via the expression:
\begin{equation}
a_{\eta^\prime p}= -\frac{1}{4\pi}\frac{M_p}{m_{\eta^\prime} + M_p} t_{\eta^\prime p \to \eta^\prime p } \ ,
\label{eq:scatlength }
\end{equation}
when only the Weinberg-Tomozawa terms obtained from the Lagrangian of Eq.~(\ref{lagran}) are included ($PB$) and incorporating also the coupling to $VB$ states ($PB-VB$). In this later case, we show the results for various approaches to the $PB-VB$ coupling: including only the box term of Fig.~\ref{fig:Box} (box) and including also the interacting terms of Fig.~\ref{fig:vectorloop} (all). For each approach, we display the results for the normal coupling obtained from diagram \ref{fig:vectorloop}(c) (normal) and adding also the contributions of the anomalous diagram \ref{fig:vectorloop}(d) (total).

   For completeness, the table also gives results for the $\eta p$ scattering length, $a_{\eta p}$, which shows to be only moderately affected by the inclusion of the $VB$ components, of which the interacting terms produce more pronounced changes.
The small differences between the ``normal" and ``total" results of the
${\eta p}$ scattering length are only possible from multiple scattering processes implicit in the Bethe-Salpeter equation, which contain intermediate $\eta^\prime N \to V B$ transitions involving the anomalous coupling.

 In the case of the $\eta^\prime p$ scattering length, we observe that the standard $PB$ model produces a very small value for it, which is connected to the fact that the Lagrangian of Eq.~(\ref{lagran}) does not give a contribution for SU(3) singlets, and the $\eta^\prime$ meson is mostly a singlet. The incorporation of the $VB$ states into the scheme, via the normal coupling, gives very moderate changes both when only the box diagram is considered and even when the interacting terms are added. The picture changes considerably when the anomalous couplings of the $ \eta^\prime$ to the vector mesons is added. Then, at the box-diagram level, the real part of $a_{\eta^\prime}$ increases by an order of magnitude, while the imaginary part  decreases slightly. The interacting terms increase further the real part but they affect especially  the imaginary part, which roughly doubles its size, the reason being the presence of the $VB$ decay channels, $K^* \Lambda \to \pi K \Lambda$ and $K^* \Sigma \to \pi K \Sigma$, which are opened at the threshold energy of the $\eta^\prime p$ system, $\sqrt{s}=1896$ MeV.

\begin{table}[htbp]
    \setlength{\tabcolsep}{0.3cm}
\begin{center}
\begin{tabular}{lc|cc|cc}
 & & \multicolumn{2}{c|}{$a_{\eta p}$} & \multicolumn{2}{c}{$a_{\eta^\prime p}$}
\\
\multicolumn{2}{c|}{ } & \multicolumn{2}{c|}{[fm]} & \multicolumn{2}{c}{[fm]}  \\
\hline
 $PB$ & &  \multicolumn{2}{c|}{$0.210 + {\rm i} 0.251$}
& \multicolumn{2}{c}{$0.0017+{\rm 0.0139}$} \\
\hline
\multirow{3}{*}{$PB-PV$}&  & (normal) & (total) & (normal) & (total) \\
 &  (box) & $0.188+ {\rm i}0.229$ &   $0.187+ {\rm i}0.232$  &
 $0.0024+{\rm i}0.0138$ &  $0.0164+{\rm i}0.0109$ \\
& (all) & $0.268+ {\rm i}0.240$ &   $0.264+ {\rm i}0.242$  &
 $0.0027+{\rm i}0.0143$ &  $0.0210+{\rm i}0.0192$  \\
\hline
\end{tabular}
\end{center}
\caption{Scattering lengths of $\eta p$ and $\eta^\prime p$ for the model that considers only $PB$ states and the model that also considers the coupling to $VB$ states ($PB-VB$). In the later case, several results are displayed: those obtained from the box-only term of Fig.~\ref{fig:Box} (box) or including also the interacting terms of Fig.~\ref{fig:vectorloop} (all), considering only the normal coupling of Fig.~\ref{fig:vectorloop}(c) (normal) and adding also the anomalous one
Fig.~\ref{fig:vectorloop}(d) (total).\label{tab:scat}}
\end{table}

\begin{figure}[ht]
\begin{center}
\includegraphics[width=0.65\textwidth]{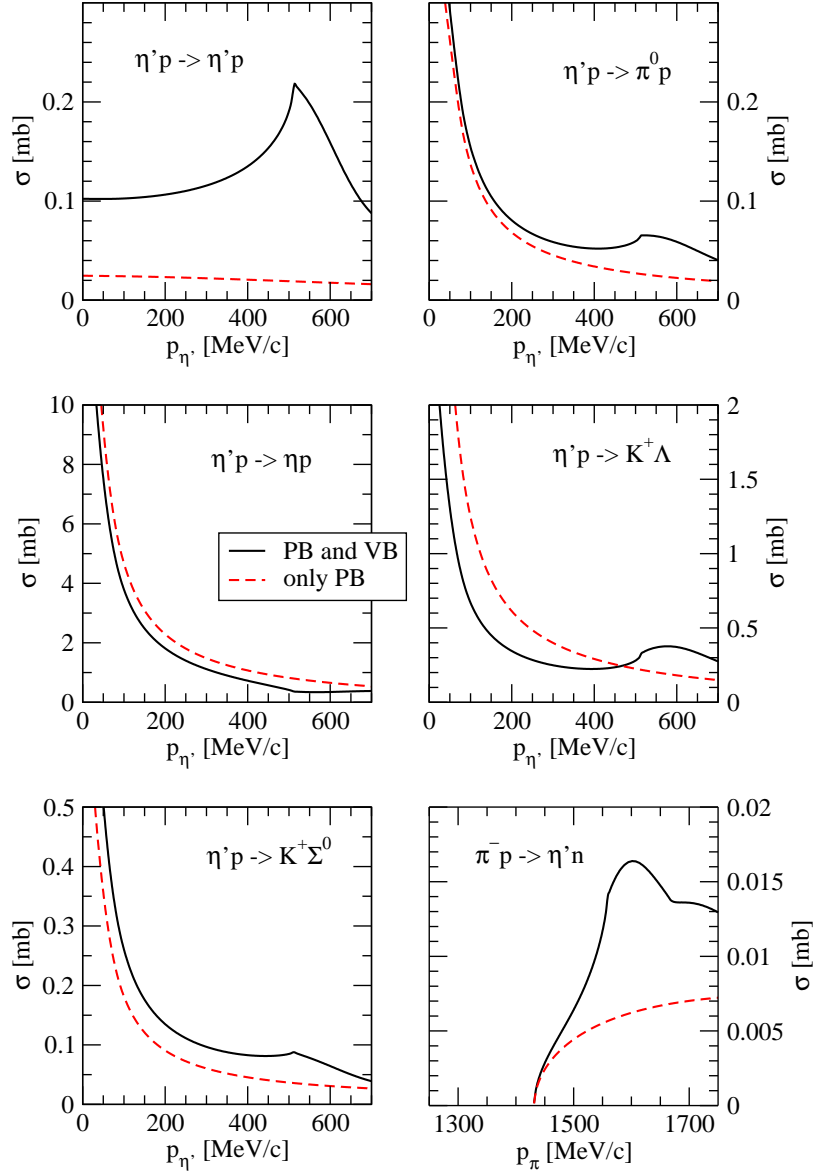}
\end{center}
\caption{Elastic and inelastic $\eta^\prime p$  cross sections, as functions of the $\eta^\prime$ momentum in the lab frame, together with the cross section of the reaction $\pi^- p \to \eta^\prime n$ as function of the pion momentum. Dashed lines correspond to the results obtained with the standard model that considers only the coupling to $PB$ states, while solid lines also contain the coupling to the $VB$ channels.}
\label{fig:xsec}
\end{figure}

With the normalization of our scattering amplitudes $t_{ij}$, the cross sections are obtained from:
\begin{equation}
\sigma_{ij}=\frac{1}{4\pi}\frac{M_i M_j}{s}\frac{k_j}{k_i}\mid t_{ij}\mid^2 \ ,
\label{eq:cross}
\end{equation}
where $M_i$, $M_j$ are the baryon masses and $k_i$, $k_j$ the c.m. momenta in the incoming and outgoing channels. In Fig.~\ref{fig:xsec}, we show the elastic $\eta^\prime p \to \eta^\prime p$ and inelastic  $\eta^\prime p \to \pi^0 p$,  $\eta^\prime p \to \eta p$,  $\eta^\prime p \to K^+ \Lambda$,  $\eta^\prime p \to K^+ \Sigma^0$ cross sections, as functions of the $\eta^\prime$ momentum in the lab frame, together with the cross section of the reaction $\pi^- p \to \eta^\prime n$, for which there is some data \cite{Rader:1973mx}, as function of the pion momentum. We show results obtained with the standard model that considers only the coupling to $PB$ states (dashed lines) and with the model that incorporates the coupling to the $VB$ channels that have a stronger influence at energies around the $\eta^\prime p$ threshold, namely $K^* \Lambda$ and $K^* \Sigma$. Depending on the process, the inclusion of the $VB$ channels may produce a positive or negative interference with the standard $PB$ terms. However,  we observe in all cases a pronounced structure around $p_{\eta^\prime} = 500 $~MeV/c, or $p_{\pi}=1600$~MeV/c, which corresponds to the energy of the resonance at 1970 MeV, generated dynamically with the model of Ref.~\cite{angelsvec} and which could be associated to some of the states in the PDG with the appropriate quantum numbers. The strongest relative enhancements of the cross sections are seen for the elastic channel and for the $\pi^- p \to \eta^\prime n$ reaction. However, in spite of these drastic changes, the results in these two cases are far from the experimental information that, although limited, we have presently available.
The obtained elastic $\eta^\prime p \to  \eta^\prime p$ amplitude produces a scattering length of $\mid a_{\eta^\prime p} \mid = 0.03$ fm, as can be derived from Table~\ref{tab:scat}, instead of the experimental estimate of $\mid a_{\eta^\prime p} \mid \sim 0.1$ fm \cite{Moskal:2000pu}. In addition, the $\pi^- p \to \eta^\prime n$ cross section at a pion lab momentum of 1600 MeV/c has been measured to be around 0.1 mb \cite{Rader:1973mx}, a value six times larger than what we find after considering the coupling to $VB$ states.

  Therefore, even after including the effect of the resonances at around 2 GeV via the coupling to $VB$ states, the model reveals insufficient. We still have one
parameter in the theory, the strength $\alpha$ of the Lagrangian coupling singlet mesons to baryons, as written in Eq.~(\ref{eq:singletcoup}). We change the value of this parameter to obtain a modulus of the $\eta^\prime N$ scattering length around 0.1 fm. In particular, we choose values $\mid a_{\eta^\prime p} \mid=0.075$, $0.1$ and $0.15$ fm. Our results are shown in Table~\ref{tab:singlet}, where we also list, for completeness, the corresponding results for the $\eta p$ scattering length.
We observe that each value of the assumed $\eta^\prime p$ scattering length admits two types of solutions for $\alpha$, one positive (repulsive) and one negative (attractive). For a given sign, the changes in  $\alpha$ modify the real part of $a_{\eta^\prime p}$ in a nearly proportional way, while the imaginary part remains practically constant. This is simply understood from the singlet and octet components of the $\eta$, $\eta^\prime$ mesons [see Eq.~(\ref{eq:mezclas})] and the structure of the singlet Lagrangian, which induces the largest effect on the $\eta^\prime N \to \eta^\prime N$ amplitude as it behaves like $\cos^2\theta_P$. The
$\eta^\prime N \to \eta  N$ amplitude will be also affected but in a more moderate way since it goes like
$\cos \theta_P \sin \theta_P$. The rest of inelastic transition amplitudes of $\eta^\prime N$ to states with pure octet mesons,  $\eta^\prime N \to K \Lambda$, $\eta^\prime N \to K \Sigma$,
and $\eta^\prime N \to \pi N$, are not affected, up to multiple scattering terms which are small in this region. Consequently, due to the optical theorem, the imaginary part of the $\eta^\prime N \to \eta^\prime N$ amplitude should not be much affected, as it is the case. This also means that our model, in spite of using the $\eta^\prime N$ scattering length as a free parameter, provides a robust prediction for the inelastic channels. We note that the magnitude of the singlet strength $\alpha$ that reproduces the experimental estimate of  $\mid a_{\eta^\prime p} \mid\sim 0.1$ fm amounts to 10-20\% of the size of the WT coefficients $C_{ij}$ obtained from the standard $PB$ Lagrangian of Eq.~(\ref{lagran}).

It is interesting to see that the two $\alpha$ solutions for a common value of  $\mid a_{\eta^\prime p} \mid$ give different values for
the $\eta p$ scattering length. However, the differences are quite moderate and cannot be used as a way of discriminating between the positive or the negative value of the singlet strength $\alpha$.

\begin{table}[htbp]
    \setlength{\tabcolsep}{0.3cm}
\begin{center}
\begin{tabular}{c|cc|cc}
$\alpha$ & $a_{\eta p}$ & $\mid a_{\eta p} \mid$   &
$a_{\eta^\prime p}$ & $\mid a_{\eta^\prime p} \mid$  \\
 & [fm] & [fm] & [fm] & [fm] \\
\hline
$-0.126$ & $0.272+{\rm i}0.246$ & $0.367$ & $0.073 +{\rm i}0.019$ & 0.075 \\
$0.204$ & $0.247+{\rm i}0.233$ & $0.340$ & $-0.072 +{\rm i}0.020$ & 0.075 \\
\hline
$-0.193$ & $0.276+{\rm i}0.248$ & $0.371$ & $0.098 +{\rm i}0.020$ & 0.1 \\
$0.256$ & $0.241+{\rm i}0.231$ & $0.334$ & $-0.098 +{\rm i}0.020$ & 0.1 \\
\hline
$-0.333$ & $0.282+{\rm i}0.251$ & $0.378$ & $0.149 +{\rm i}0.020$ & 0.15 \\
$0.352$ & $0.228+{\rm i}0.225$ & $0.320$ & $-0.149 +{\rm i}0.021$ & 0.15 \\
\hline
\end{tabular}
\end{center}
\caption{The negative and positive values of the singlet strength $\alpha$ that produce $\eta^\prime p$ scattering lengths of 0.075, 0.1 and 0.15 fm. The corresponding values of the $\eta p$ scattering lengths are also given. \label{tab:singlet}}
\end{table}

\begin{figure}[ht]
\begin{center}
\includegraphics[width=0.65\textwidth]{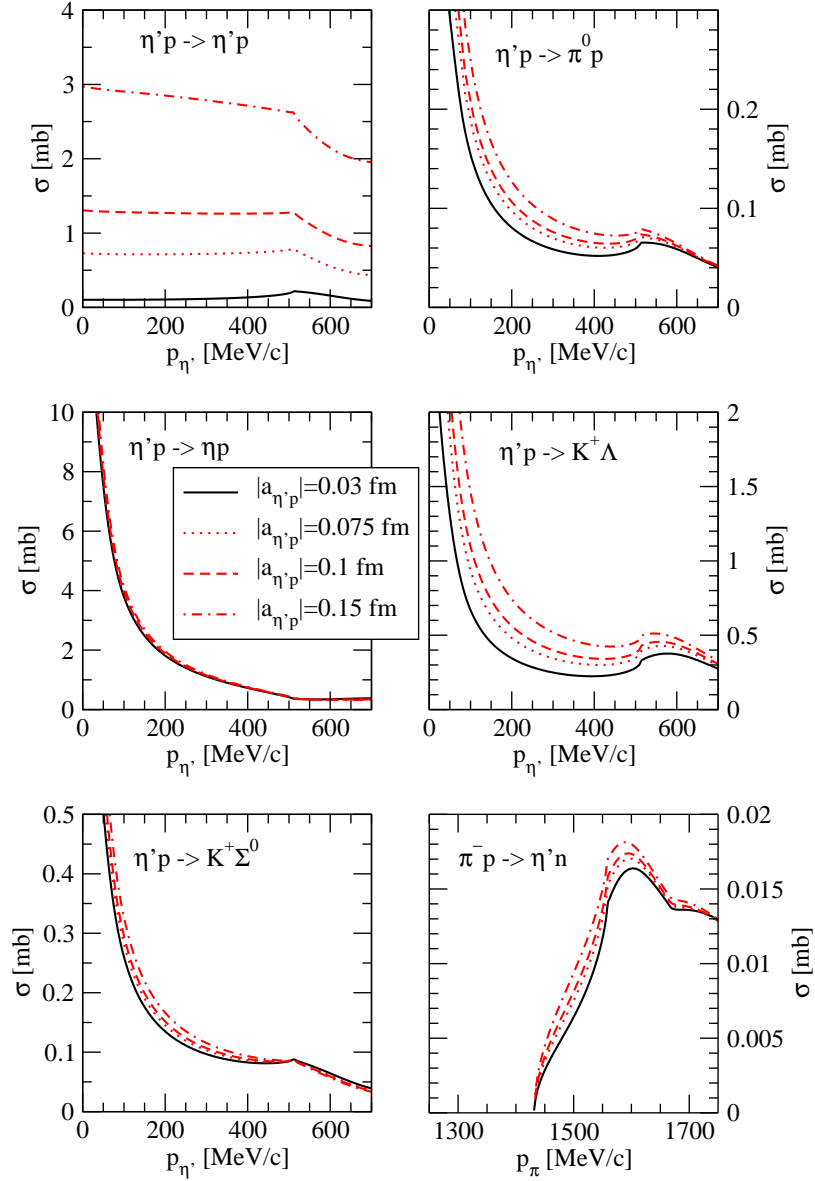}
\end{center}
\caption{Elastic and inelastic $\eta^\prime p$  cross sections, together with the cross section of the reaction $\pi^- p \to \eta^\prime n$, for various models giving rise to different values of $\mid a_{\eta^\prime p}\mid$.}
\label{fig:xsec_singlet}
\end{figure}

In Fig.~\ref{fig:xsec_singlet}  we show the elastic and inelastic $\eta^\prime p$  cross sections, together with the cross section of the reaction $\pi^- p \to \eta^\prime n$, for various models corresponding to different values of $\mid a_{\eta^\prime p}\mid$, obtained with negative values of $\alpha$. Similar results would be obtained for the equivalent positive values of $\alpha$. The elastic $\eta^\prime p$ cross section experiences a drastic enhancement with increasing values of $\mid a_{\eta^\prime p}\mid$. As expected, the changes in the inelastic cross sections are more moderate, although in some cases the relative increase of the cross section may become significant.

\begin{figure}[ht]
\begin{minipage}[t]{0.45\linewidth}
\centering
\includegraphics[width=\textwidth]{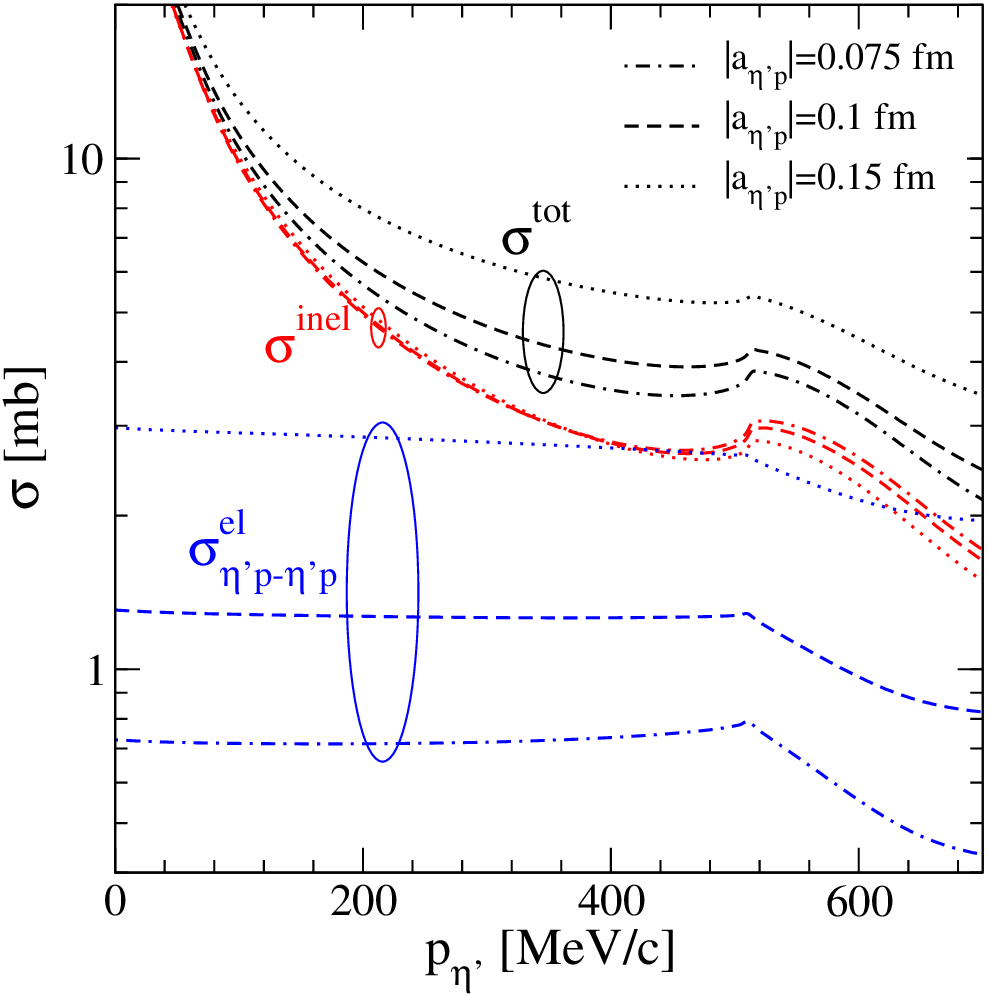}
\caption{Elastic, inelastic and total $\eta^\prime N$ cross sections as functions of the $\eta^\prime$ momentum in the lab frame, for different values of $\mid a_{\eta^\prime p}\mid$.}
\label{fig:xcross_tot_el_inel}
\end{minipage}
\hspace{0.5cm}
\begin{minipage}[t]{0.45\linewidth}
\centering
\includegraphics[width=\textwidth]{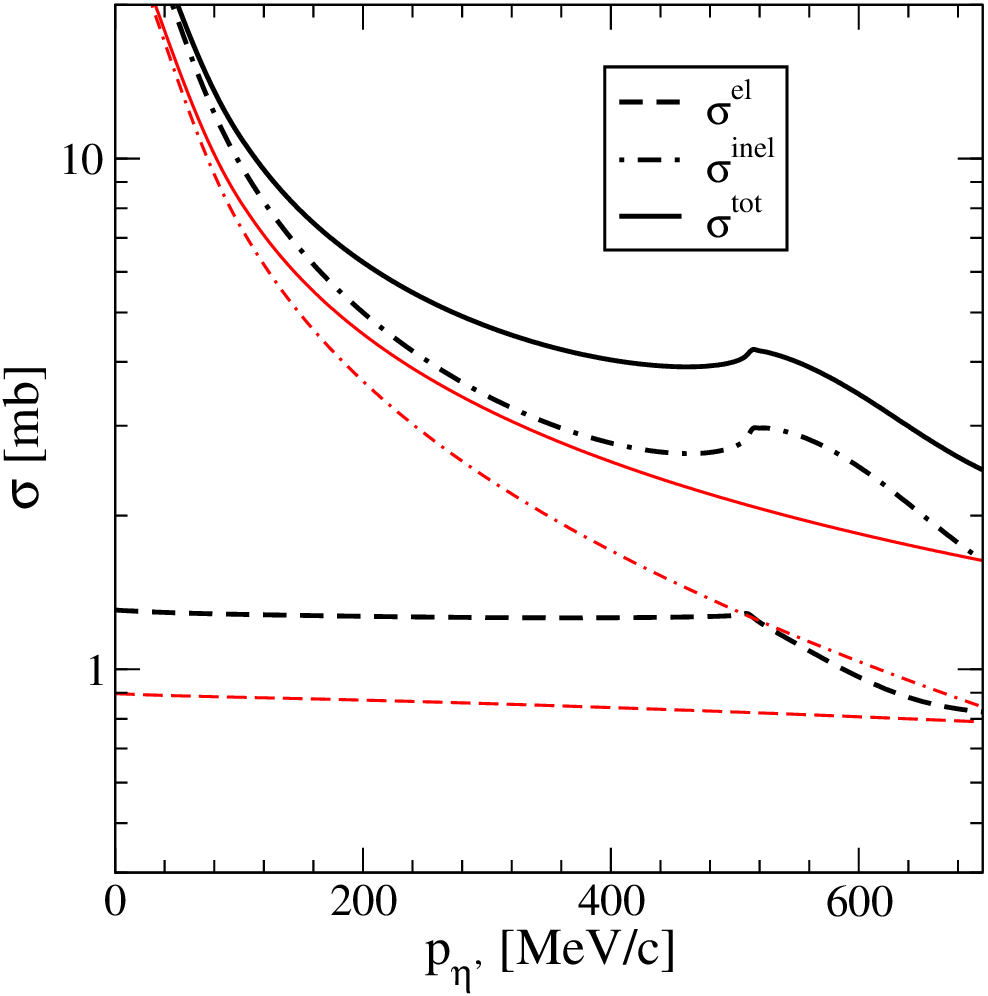}
\caption{Elastic, inelastic and total $\eta^\prime N$ cross sections as functions of the $\eta^\prime$ momentum in the lab frame, for $\mid a_{\eta^\prime p}\mid=0.1$ fm, with (thick lines) or without (thin lines) the coupling to the $VB$ channels.}
\label{fig:xcross_noVB}
\end{minipage}
\end{figure}

The total $\eta^\prime p$ cross section can be derived from the elastic scattering amplitude via the optical theorem:
\begin{equation}
{\rm Im}\, t_{\eta^\prime p \to \eta^\prime p}= -\frac{p_{\rm c.m.} \sqrt{s}}{M_p}\sigma_{\rm tot} \ ,
\label{eq:optical}
\end{equation}
where $p_{\rm c.m.}$ is the $\eta^\prime$ momentum in the c.m. frame. By subtracting the elastic cross section, we obtain the inelastic one, $\sigma^{\rm inel}=\sigma^{\rm tot}-\sigma_{\eta^\prime p \to \eta^\prime p}^{\rm el}$. The elastic, inelastic and total $\eta^\prime N$ cross sections are displayed
in Fig.~\ref{fig:xcross_tot_el_inel} for different values of $\mid a_{\eta^\prime p}\mid$. As mentioned above, the more drastic changes are seen in the elastic cross-section, the inelastic one being much less affected and becoming, in this way, a genuine prediction of our model.
We observe that the inelastic cross sections to the pseudoscalar-baryon channels displayed in Fig.~\ref{fig:xsec_singlet} do not provide the amount of inelastic cross section of Fig.~\ref{fig:xcross_tot_el_inel}. For $\mid a_{\eta^\prime p}\mid=0.1$ fm, for instance, and a value of 200~MeV/c for the $\eta^\prime$ momentum, the inelastic cross sections of Fig.~\ref{fig:xsec_singlet}, multiplying $\eta^\prime p \to \pi^0 p$ and $\eta^\prime p \to K^+ \Sigma^0$ by 3 to account for all isospin channels, add up to about 3.5 mb, while $\sigma^{\rm inel}\sim 5$ mb in
Fig.~\ref{fig:xcross_tot_el_inel}. The difference is due to the inelasticities coming from the $VB$ channels $K^* \Lambda$ and $K^*\Sigma$ included in our model, an effect that is more clearly seen in Fig.~\ref{fig:xcross_noVB}, where the elastic (dashed lines), inelastic (dash-dotted lines) and total (solid lines) cross sections, obtained with the value of $\alpha$ that gives $\mid a_{\eta^\prime p}\mid=0.1$ fm, are shown for the full model (thick lines) or when the coupling to the $VB$ channels is switched off (thin lines). Indeed, we observe a large enhancement in the inelastic cross section produced by the $VB$ channels, especially close to the position of the $K^*\Sigma$ resonance which is visible around 500 MeV/c of $\eta^\prime$ momentum. We have also studied the amount of inelasticity associated to the intermediate $\rho N$ channels and have found it to be negligible.

     There is one source of information for the inelastic cross section, coming from the recent experiment
 on the transparency ratio of $\eta^\prime $ photoproduction in nuclei \cite{nanova}. So far, the preliminar analysis of the data give an estimate of 10-12 mb for the $\eta^\prime N$ inelastic cross section around $p_{\eta^\prime}= 900$ MeV/c. This estimate is an upper bound for the one-body absorption inelasticities, since in the nucleus the absorption of the $\eta^\prime $ can also proceed via two nucleons, which would add to the inelastic $\eta^\prime $ reactions with a single nucleon. In any case, the $\eta^\prime $ energies of the transparency ratio experiments are far too high for our model, the validity of which we trust for momenta smaller than $p_{\eta^\prime}= 600$ MeV/c, implying kinetic energies for the  $\eta^\prime $ smaller than 200 MeV. Data analyses with momentum cuts will be available in the future and, although the statistics will then be reduced, they will help in constraining the properties of the $\eta^\prime N $ interaction.

\section{Conclusions}

   We have obtained the $\eta^\prime N$ scattering amplitude within a coupled channels chiral unitary approach, using the standard $\eta_1-\eta_8$ mixing for the decomposition of the physical $\eta$, $\eta^\prime$ states into the singlet and octet fields of SU(3). The particular form of the lowest order chiral Lagrangian makes the contribution of the singlet vanish.  Consequently, since the $\eta^\prime$ is mostly a singlet, the obtained $\eta^\prime N$ amplitude  is very small and produces a value for the scattering length which disagrees with recent experimental analysis that estimate it to be around 0.1 fm. Moreover, the fact that the calculated $\pi^- p \to \eta^\prime n$ cross section is also extremely small compared with experiment indicates that something important is missing in the approach.

   We have found that the coupling of the $PB$ channels to the $VB$ ones is very significant because, in the threshold region of $\eta^\prime N$ states, the $VB$ interaction leads to a dynamically generated resonance with the same quantum numbers as the $\eta^\prime N$ system in s-wave. We have developed the formalism to connect the $PB$ with the $VB$ states, including anomalous $VVP$ couplings which turn out to be very important for the $\eta^\prime N\to VB$ transitions. The inclusion of the $VB$ channels induces important changes in the  $\eta^\prime N$  reactions but still leads to a small scattering length compared to the experimental estimate.

   Finally, we have introduced a Lagrangian that couples the baryons to the singlet SU(3) meson and have made changes in the unknown strength in order to get acceptable values of the $\eta^\prime N$ scattering length. Yet, this does not change much the inelastic cross sections, which become then robust predictions of our theory. There is not much information to compare our results for the inelastic cross sections, except for recent results on the transparency ratio obtained from $\eta^\prime$ photoproduction reactions in nuclei. The experimental results offer an upper limit of the inelastic cross section and, at present, they are provided for values of the $\eta^\prime$ momenta too high for our model. Future measurements with momentum cuts will allow us a proper comparison with data and hopefully gain further information on the $\eta^\prime N $ interactions.

\section*{Acknowledgments}
We would like to thank C. Hanhart and J. Taron for useful information and J Garzon for valuable checks.
This work is partly supported by  projects FIS2006-03438, FIS2008-01661 from the Ministerio de Ciencia e Innovaci\'on (Spain), by the Generalitat Valenciana in the program Prometeo and
 by the Ge\-ne\-ra\-li\-tat de Catalunya contract 2009SGR-1289.
 This research is part of the European
 Community-Research Infrastructure Integrating Activity ``Study of
 Strongly Interacting Matter'' (acronym HadronPhysics2, Grant
 Agreement n. 227431) and of the EU Human Resources and Mobility
 Activity ``FLAVIAnet'' (contract number MRTN--CT--2006--035482),
 under the Seventh Framework Programme of EU.

\end{document}